\documentclass[11pt]{article}
\pdfoutput=1
\usepackage{amsmath}
\usepackage{amssymb}
\usepackage{graphicx,bbm,mathrsfs}
\usepackage{nicefrac}
\usepackage{slashed}
\usepackage{bbm}
\usepackage{geometry}
\usepackage{empheq}

\usepackage{jheppub}

\usepackage{mathtools}

\usepackage{dcolumn}   
\usepackage{bm}        
\usepackage{graphicx,mathrsfs}
\usepackage{nicefrac}
\usepackage{multirow}
\usepackage{color}
\usepackage{mathtools}
\usepackage{makecell}

\usepackage{environ} 
\usepackage{lipsum} 
 \NewEnviron{Smaller11}{
           \scalebox{1.1}{$\BODY$} 
 } 
 \NewEnviron{Smaller08}{
           \scalebox{0.8}{$\BODY$} 
 }

\newcommand{\be}{\begin{equation}}
\newcommand{\ee}{\end{equation}}
\newcommand{\bea}{\begin{eqnarray}}
\newcommand{\eea}{\end{eqnarray}}

\newcommand{\Lv}{{\bf L}}

\newcommand{\F}{{ \bm F}}
\newcommand{\x}{{\bm x}}

\newcommand{\bpartial}{{\bm \partial}}
\newcommand{\W}{W}

\renewcommand{\a}{\xi}

\usepackage{wasysym}
\usepackage{hhline,colortbl}
\usepackage{xcolor}

\newcommand{\nn}{\nonumber}

\mathchardef\mhyphen="2D

\newcommand{\cB}{\star}

\usepackage{titlesec}

\titleformat*{\section}{\Large\bfseries}
\titleformat*{\subsection}{\large\bfseries}
\titleformat*{\subsubsection}{\large\bfseries}
\titleformat*{\paragraph}{\large\bfseries}
\titleformat*{\subparagraph}{\large\bfseries}

\makeatletter
\newcommand*{\prodsym}{%
  \DOTSB
  \mathop{
    \mathchoice
      {\rlap{\kern.3em\rotatebox[origin=c]{-90}{}}{\prod}}
      {\vcenter{\rlap{\kern.2em\rotatebox[origin=c]{-90}{}}}{\prod}}
      {\sum}{\sum}
  }\slimits@
}
\makeatother

\makeatletter
\DeclareFontFamily{OMX}{MnSymbolE}{}
\DeclareSymbolFont{MnLargeSymbols}{OMX}{MnSymbolE}{m}{n}
\SetSymbolFont{MnLargeSymbols}{bold}{OMX}{MnSymbolE}{b}{n}
\DeclareFontShape{OMX}{MnSymbolE}{m}{n}{
    <-6>  MnSymbolE5
   <6-7>  MnSymbolE6
   <7-8>  MnSymbolE7
   <8-9>  MnSymbolE8
   <9-10> MnSymbolE9
  <10-12> MnSymbolE10
  <12->   MnSymbolE12
}{}
\DeclareFontShape{OMX}{MnSymbolE}{b}{n}{
    <-6>  MnSymbolE-Bold5
   <6-7>  MnSymbolE-Bold6
   <7-8>  MnSymbolE-Bold7
   <8-9>  MnSymbolE-Bold8
   <9-10> MnSymbolE-Bold9
  <10-12> MnSymbolE-Bold10
  <12->   MnSymbolE-Bold12
}{}

\let\llangle\@undefined
\let\rrangle\@undefined
\DeclareMathDelimiter{\llangle}{\mathopen}%
                     {MnLargeSymbols}{'164}{MnLargeSymbols}{'164}
\DeclareMathDelimiter{\rrangle}{\mathclose}%
                     {MnLargeSymbols}{'171}{MnLargeSymbols}{'171}
\makeatother

\begin{document}

\vspace*{4mm}

\begin{center}

\begin{minipage}{20cm}
\begin{center}
\hspace{-5cm }
\huge
\sc
Casimir Forces  in CFT with Defects \\
\hspace{-5cm }
 and Boundaries
\end{center}
\end{minipage}
\\[30mm]


\renewcommand{\thefootnote}{\fnsymbol{footnote}}

{\large  
Philippe~Brax~$^a$\footnote{phbrax@gmail.com} and Sylvain~Fichet~$^{b}$\footnote{sylvain.fichet@gmail.com}\,
}\\[8mm]

\end{center}

\noindent \quad\quad\quad $^a$ \textit{Institut de Physique Th\'{e}orique, Universit\'e Paris-Saclay, CEA, CNRS, }

\noindent \quad\quad\quad\quad \textit{ F-91191 Gif/Yvette Cedex, France}
\\

\noindent \quad\quad\quad $^{b}$\textit{ Centro de Ciencias Naturais e Humanas, Universidade Federal do ABC,}

\noindent \quad\quad\quad\quad \textit{Santo Andre, 09210-580 SP, Brazil}

\addtocounter{footnote}{-2}

\vspace*{12mm}

\begin{center}
{  \bf  Abstract }
\end{center}

We investigate the quantum forces occurring between the  defects and/or boundaries of a conformal field theory (CFT). 
We   propose to model imperfect defects and boundaries 
as localized relevant double-trace  operators that deform the CFT. 
 Our focus is on pointlike and codimension-one  planar defects.
In the case of two parallel membranes, we point out that the CFT 2-point function tends to get confined and develops a tower of resonances with constant decay rate when the operator dimension approaches the free field dimension. 
Using a functional formalism, we  compute the quantum forces induced by the CFT between a variety of configurations of pointlike defects, infinite plates and membranes. Consistency arguments imply that these quantum forces are attractive at any distance. 
Forces of Casimir-Polder type appear in the UV, while forces of Casimir type appear in the IR, in which case the CFT gets repelled from the defects. Most of the forces behave as a non-integer power of the separation, controlled by the  dimension of the double-trace deformation. In the Casimir regime of the membrane-membrane configuration, the quantum pressure  behaves universally as $1/\ell^d$, however information about the double-trace nature of the defects still remains encoded in the  strength of the pressure.

 \newpage
\setcounter{tocdepth}{2}
\tableofcontents

\section{Introduction}

Quantum Field Theory (QFT) predicts that  macroscopic bodies can experience 
forces of purely quantum nature \,\cite{CP_original,Casimir_original}.  Such quantum forces are usually computed within the framework of weakly coupled QFT, see e.g. \cite{Milton:2004ya, Klimchitskaya:2009cw, Woods:2015pla, Bimonte:2017bir, Bimonte:2022een, bordag2009advances} for modern reviews. 
In this paper, we propose to explore the quantum forces that arise in a particular class of QFTs in which calculations are possible even at strong coupling:  Conformal Field Theories (CFTs).

Conformal field theories are ubiquitous in the real world. Many thermodynamic and quantum critical points exhibit conformal invariance.
For example, the liquid-vapor critical points, the superfluid transition in liquid helium, Heisenberg magnets,  are all described by the  same family of scalar 3D CFTs, see e.g. \cite{Poland:2022qrs,Dantchev:2022hvy}\,.
CFTs are also ubiquitous in the space of quantum field theories: most  Renormalisation Group (RG) flow end on a CFT, either in the IR or the UV. Reversing the logic, one can also think of generic weakly coupled QFTs as CFTs deformed by operators that are either relevant or irrelevant. 

The CFTs that appear in  the real world are not perfect.
Critical systems obtained in the laboratory  certainly have boundaries. Moreover, real-world CFTs can contain impurities of various codimensions.
A subfield of CFT studies focuses on extracting data from CFTs with boundaries and defects using inputs from symmetry, unitarity and causality.\,\footnote{\label{footnote1} See e.g.  \cite{McAvity:1995zd, Liendo:2012hy, Gaiotto:2013nva, Gliozzi:2015qsa, Billo:2016cpy} for some seminal papers,  \cite{Behan:2020nsf,  Behan:2021tcn,Barrat:2022psm,Bianchi:2022ppi, Gimenez-Grau:2022czc,Antunes:2022vtb,Bianchi:2022sbz,Gimenez-Grau:2022ebb,
SoderbergRousu:2023nvd,Barrat:2023ivo,Cuomo:2023qvp,
Behan:2023ile} for recent progresses, \cite{Andrei:2018die,  herzog_lecture} for recent reviews. } 
{The present work does not pursue this  approach. 
Our focus  is rather  a set of observable phenomena that  we compute  via QFT methods  adapted to the CFT context. }

Boundaries and defects in the real world are not perfect either.  Physical defects  cannot, in general, be thought as perfect truncations of the spatial support of a field theory with fluctuations of any wavelength.
A more realistic description of defects should feature some notion of smoothness. 
The modeling of such imperfect defects and boundaries is somewhat familiar from weakly coupled QFT. There, a defect is sometimes modelled by a bilinear operator, whose spatial support represents  the defect  \,\cite{Graham:2002fw,Graham:2002xq,Graham:2003ib}.  Within such a model, the defect perfectly repels the field only asymptotically in the infrared. More generally, for arbitrary wavelengths, the quantum field propagates to some extent inside the defect \cite{Brax:2018grq,Brax:2022wrt}.  One of the aims of this paper  is to  model  imperfect defects in CFTs in an analogous manner. This is done in section \ref{se:DTD}. 

 The second aim of this paper is the computation of observable quantities: the quantum forces  induced by the CFT between pairs of defects and/or boundaries.
{We assume that spacetime dimension is equal or larger than $3$, see e.g. \cite{Cardy:1986gw,Kleban_1991,Eisen2004,Bimonte2013,Geng:2021iyq} for  Casimir-type computations in 2d CFT. 
We mainly focus on quantum fluctuations in spacetime, however our approach can analogously applies to thermal fluctuations in Euclidian space --- since  quantum  and statistical field theories are related via Wick rotation. In the thermodynamic context,  the  fluctuating field describes  an order parameter of a continuous phase transition. 
One usually uses the term  {\it critical} Casimir forces \cite{Dantchev:2022hvy} to refer to forces appearing near criticality, where the system becomes a CFT.   
The quantities  computed in the thermal case are however slightly different from the ones  in QFT. In QFT one computes a force or  potential between non-relativistic bodies, while in the thermal case one typically computes the free energy at criticality. 
}

 Our results {on quantum forces} are presented in  section  \ref{se:Casimir}, {where we also discuss monotonicity and the connection to critical Casimir forces. }
{In the process} we analyze the properties of  two-point correlators confined between membranes  in section \ref{se:membranes}.  
 Section \ref{se:basics} contains the necessary introductory material
 and section \ref{se:summary} contains  a summary of our results.


\section{Basics}
\label{se:basics}

\subsection{CFT Rudiments }
\label{se:CFT}

A Conformal Field Theory is a field theory that is invariant under the Conformal Group $SO(d,2)$ --- or $SO(d+1,1)$ in Euclidian space. CFTs are  fairly miraculous.
The symmetries of the conformal group are so strong that they fully constrain both the 2-point and 3-point correlation functions of any operator. 
Still due to symmetries, operators and states are in one-to-one correspondence and the operator product expansion (OPE) has a finite  radius of convergence. 
 The OPE, combined with crossing symmetry, provides nontrivial constraints on 4-point correlators, which is the theme of the  ``Conformal Bootstrap'' program. See \cite{Rychkov:2016iqz,Simmons-Duffin:2016gjk,Poland:2018epd,Chester:2019wfx,Qualls:2015qjb,Gillioz:2022yze} for modern reviews and lecture notes  on CFTs.  
In this paper, we only need the most basic features of CFTs, and no prior CFT knowledge is needed.  

The symmetries of the conformal group impose that so-called primary operators ${\cal O}_i$  have 2-point correlators of the form
\be\langle {\cal O}_i(x_1) {\cal O}_j(x_2)\rangle =\frac{ a_i \delta_{ij}}{ x_{12}^{2\Delta_i} }\ee 
with $x^2_{12}=(x_1-x_2)^\mu (x_1-x_2)_\mu$. $\Delta_i$ is the scaling dimension of ${\cal O}_i$ under the dilatation operator.  

The overall constant $a_i$ is not fixed by symmetries. In this work we adopt the normalization $a_i\equiv 1$. CFT unitarity implies that an operator is a free field if and only if $\Delta=\frac{d-2}{2}$. For a canonically normalized 4D free field, we would have $a_i \to \frac{1}{4\pi^2}$.  We  convert to this normalization when comparing with 4D free field results throughout this paper.

The formal CFT operators $O_i$ can be understood as traces of combinations of matrices, such as the irreducible representations  of an internal $SU(N)$ group. This is why operators of the form $[{\cal O}(x)]^n$ are usually called $n$-uple trace operators. In this paper, a central role is played by the double-trace operators $[{\cal O}(x)]^2$.
An operator is said to be relevant, marginal and irrelevant if $\Delta<d$, $\Delta=d$, and $\Delta>d$ respectively.

We further assume that the CFT has a large number of degrees of freedom, i.e. large $N$, such that 't\,Hooft's large-$N$ expansion applies. This assumption renders many calculations possible, here we only need to work at leading order of the large $N$ expansion.\,\footnote{{Moreover we only focus on 2-pt correlators. At large $N$ the 2-pt correlators that we compute amount to those of a scalar generalized free field (GFF), i.e.  a free scalar  with dimension $\Delta> \frac{d-2}{2}$ \cite{Greenberg:1961mr}.  
An actual GFF would appear by taking $N\to \infty$, in which case all the higher-point correlators of a GFF are trivially expressed as a function of the 2-pt GFF correlator via Wick's theorem, see e.g.  \cite{Kap:lecture}. 
In this work  we do not need to take infinite $N$, which is known to be an ill-defined limit in CFT and beyond, see e.g. \cite{Dymarsky:2014zja,Fichet:2022ixi}. We assume large but finite $N$,  and  all our results are given up to $O(\frac{1}{N^2})$ corrections. } } In this regime, the scaling dimension of the double-trace operator is simply $\Delta_{{\cal O}^2}=2\Delta+O(\frac{1}{N^2})$. 

CFTs in the real world live in finite volumes with boundaries. Furthermore they may contain impurities. This has triggered a  formal program of  studies  constraining CFTs with boundaries and defects --- the Boundary Conformal Boostrap. See footnote \ref{footnote1} for general references. 
In this note, we do \textit{not} use bootstrap techniques. 
It might be fruitful to apply bootstrap techniques to the class of defects and boundaries that we introduce further below, this is left for future work.

\subsubsection{Momentum space}

We compute the CFT 2-point function of a scalar primary $\cal O$ in momentum space  $(p^M)$. 
The Fourier transform convention is $ {\cal O} (x)=\int \frac{d^{d} p}{(2\pi)^{d}}{\cal O} (p) e^{-i p\cdot x} \label{eq:FT} \,.$ We introduce the reduced correlator
\be
\langle {\cal O}(p_1) {\cal O}(p_2) \rangle = (2\pi)^d \delta^{(d)}(p_1+p_2)
\llangle {\cal O}(p_1) {\cal O}(p_2) \rrangle \,.
\ee
We have 
$
\langle  {\cal O}(x_1)  {\cal O}(x_2) \rangle = 
\int \frac{d^dp}{(2\pi)^d} e^{-ip\cdot x_{12}} \llangle {\cal O}(p) {\cal O}(-p)\rrangle
$
and obtain 
\be
\llangle  {\cal O}(p)  {\cal O}(-p) \rrangle = - i \frac{\pi^{d/2} \Gamma(d/2-\Delta)}{\Gamma(\Delta)}\left(\frac{4}{-p^2}\right)^{d/2-\Delta}
\label{eq:2pt_momentum}\,.
\ee
A convenient way to compute the Fourier transform is via the Schwinger parametrization, see App.\,\ref{app:twopoint}.

\subsubsection{Momentum-position space}

Since we are interested in codimension-one defects, it is also useful to single out one of the  spatial dimensions corresponding to the orthogonal direction to the defects, $x^M=(y^\mu,z)$. We compute the CFT correlator in mixed position-momentum space $(p^\mu,z)$. 
For this, we
introduce the reduced mixed-space correlator
\be
\langle {\cal O}(p_1,z_1) {\cal O}(p_2,z_2) \rangle = (2\pi)^{d-1} \delta^{(d-1)}(p_1+p_2)
\llangle {\cal O}(p_1,z_1) {\cal O}(p_2,z_2) \rrangle \,.
\ee
We have 
$
\langle  {\cal O}(x_1)  {\cal O}(x_2) \rangle = 
\int \frac{d^{d-1}p}{(2\pi)^d} \llangle {\cal O}(p,z_1) {\cal O}(-p,z_2)\rrangle e^{-ip\cdot y_{12}} 
$
and obtain 
\be
\llangle  {\cal O}(p,z_1)  {\cal O}(-p,z_2) \rrangle = 
-i \frac{2\pi^{\frac{d-1}{2}}}{\Gamma(\Delta)}\left(\frac{4z_{12}^2}{-p^2}\right)^{\frac{d-1-2\Delta}{4}} K_{\frac{d-1}{2}-\Delta}\left(\sqrt{-p^2 z_{12}^2}\right) \label{eq:2pt_mixed}
\ee
that, again, can be obtained using the  Schwinger parametrization, see App.\,\ref{app:twopoint}. $K_\alpha$ is the modified Bessel function of the second kind of order $\alpha$. A useful integral representation is
\be
K_\alpha(  z)= \frac{1}{2}\left(\frac{ 2 }{z}\right)^{\alpha} \int_0^\infty \frac{dt}{t} t^\alpha e^{-t-\frac{z^2}{4t}}\,.
\label{eq:Krep}
\ee

We further introduce
\be
 \llangle  {\cal O}(p,z_1)  {\cal O}(-p,z_2) \rrangle \equiv i G(p;z_1,z_2)  \,.
\ee
With this definition, $G(p;z_1,z_2)$ is real for spacelike momenta $(p^2<0)$ or if one Wick-rotates $p$ to Euclidean space.

\subsection{Casimir Forces in the Functional Formalism}
\label{se:func}

In this note our interest lies in computing Casimir and Casimir-type forces between  defects and/or boundaries of a CFT. To this end, we use a variational approach introduced long ago in e.g. \cite{Schwinger:1977pa}, and recently exploited/developed in \cite{Brax:2022wrt}. Ref.\,\cite{Franchino-Vinas:2021lbl} used a similar approach, see also 
\cite{Bimonte:2021maf} for related developments.

In this formalism, one considers the generating functional of the correlators of the system (i.e. the free energy in Euclidean space) in the presence of a static source $J(\x)$,
\be
E[J]= i T\log Z[J]\,,\quad\quad 
Z[J]=\int {\cal D} \Phi e^{ i S[\Phi,J] }
\ee
where $\Phi$ refers collectively to the set of quantum fields. 
The  quantity $E[J]$ can be referred to as the vacuum energy evaluated in the presence of the source $J$. In the present work, the source is ultimately identified with the defects and/or boundaries of the system. 

A variation of the source produces a variation in the vacuum energy. This variation in energy is identified as a quantum version of the notion of work. We write this quantum work as
\be
\W_\lambda = -\partial_\lambda E[J_\lambda]\, \label{eq:W_def}
\ee
where $\lambda$ is a  deformation parameter. 
 In cases where the deformation of the source is simple enough, the quantum work can be  factored out as  displacement times force.  The  force that emerges from $W_\lambda$  encodes all the effects of the quantum fluctuations. This is how we compute quantum  forces in this note.

The functional formalism sketched above applies, by definition, to any field theory (either weakly or strongly coupled), and admits any kind of deformation. 
While the principle of the approach is conceptually simple, the precise formulation is slightly technical due to the fact that one needs to parametrize a generic deformation of the source. 
Assuming for simplicity that the  density is constant in $\lambda$ and $\x$ i.e. that the source is  incompressible and homogeneous, the source is written as $J_\lambda(\x)\equiv n {\bm 1}_J(\x) \equiv n \,\Theta[l_\lambda(\x)]$ with the support function $l_\lambda(\x)>0$ on the support of $J$, $l_\lambda(\x)=0$ at its boundary, and negative otherwise.\,\footnote{The general case including compressible, heterogenous sources is presented in \cite{Brax:2022wrt}.  } 
The deformation of $J_\lambda$ is described by a vector field $\Lv$ referred to as the deformation flow, such that
\be l_{\lambda+d\lambda}(\x)=l_\lambda(\x- \Lv(\x) d\lambda)\,.\ee 
Defining $\frac{\partial}{\partial \lambda}\equiv\partial_\lambda$, we obtain the definition of the quantum work as a variation in $\lambda$, written in Eq.\,\eqref{eq:W_def}. 

If  the fields couple bilinearly to the source, 
\be
S[\Phi,J]=\int dx^d \left({\cal L}[\Phi(x)]-  \frac{\a}{2} \Phi^2(x) J(\x) \right)\,,
\ee
then the quantum work is found to be \cite{Brax:2022wrt}
\be
W_\lambda = 
  -\frac{\a}{2} \int d^{d-1} \x \left\langle  \Phi(x) \Phi(x) \right\rangle_J \partial_\lambda J_\lambda(\x) \label{eq:Wdefgen} \,.
\ee
Here $\left\langle  \Phi(x) \Phi(x) \right\rangle_J$ is the two-point correlator of $\Phi$ evaluted in the presence of the $J$ source and taken at coincident point. 
This is the general formula we use in this work.
When the deformation is simple enough, the quantum work can be written as $W_\lambda=\Lv \cdot \F$ where $\F$ is identified as the quantum force.

A crucial feature highlighted by the quantum work formalism is that the matter in the source must be conserved \cite{Brax:2022wrt}. Otherwise, unphysical divergences would appear in the quantum work,  while it must be finite  by definition. 
 At constant  density, i.e. for an incompressible homogeneous source, the statement of matter conservation becomes that the deformation flow must be divergence-less, $\bpartial \cdot L(\x) =0$. This is a concrete condition that constrains the admissible deformations of $J$. 
{ An example of arbitrary deformation of an arbitrary source is
 \be
	\includegraphics[width=0.7\linewidth,trim={5cm 8.4cm 3cm 6cm},clip]{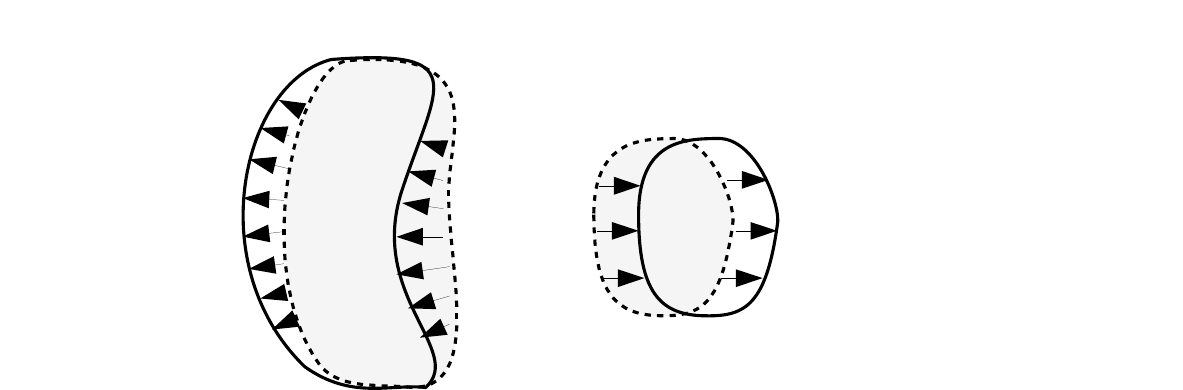}
 \nn
 \ee
where the arrows represent the divergence-less deformation flow. }

\section{Double-trace Deformations as Defects and Boundaries }
\label{se:DTD}

\subsection{Modeling Imperfect Defects and Boundaries}

In weakly coupled QFTs, it is common to model an imperfect boundary using a  mass term localized in space, $ J(\x)=m^2 {\bf 1}_J(\x)$. This mass term  dresses the $\phi$ propagator, forming a Born series 
$G_\phi(x_1,x_2) -i \int d^dx G_\phi(x_1,x)  J(\x) G_\phi(x,x_2) +\ldots $ \footnote{
The Born series can be derived by integrating out  the $\phi$ field in the partition function of the theory.
} 
In the $m^2\to\infty $ limit, the $\phi$ field is repelled from the  support of $J$, and thus acquires a Dirichlet boundary condition on $\partial J$. This can be shown  at the level of the equation of motion \cite{Brax:2022wrt}, or by inspecting the dressing of  the propagator as done further below.

The mass term is, in any $d$, a relevant operator. {Accordingly, the $m^2\to\infty $ limit can be understood as the limit of low-momentum, i.e. the infrared  regime of the RG flow.} 
With this viewpoint, one deduces  that the field is  repelled from $J$ at long distance while it propagates to some extent inside $J$ at short distance. 
This provides a simple, intuitive picture of an imperfect defect/boundary in weakly coupled QFT. {We now define a model that  reproduces such a  behaviour in CFT. }

The natural CFT analogue of a mass term is the CFT double-trace deformation. A double-trace deformation can be simply thought as a  term added to the CFT action,
\be
S^{\rm CFT}_{\rm deformed}= S^{\rm CFT} - \frac{\a}{2} \int dx^d {\cal O}^2(x)J(\x)\,.
\label{eq:DTdef}
\ee
 The deformation breaks the conformal symmetry unless $\Delta_{{\cal O}^2}=d$ exactly. Still, {in the large-$N$} limit we can compute the correlator of the deformed 2-point CFT by  dressing  the correlator in the absence of defect.

Following the intuition from weakly coupled QFT, we require that the ${\cal O}^2$ be \textit{relevant}.   
At leading order in the large-$N$ limit, this implies that the dimension of $\cal O$ must satisfy 
\be\frac{d-2}{2}\leq\Delta < \frac{d}{2}\,. \label{eq:bound}\ee
Below we further motivate this bound.

Let us first review the effect of a double-trace operator occupying the whole space. 
In that case, $J=1$. The  CFT two-point correlator is easily expressed
in momentum space,\,\footnote{{At leading order in the large-$N$ limit, the  leading effect in the dressing comes from insertions of the $-i\xi$ vertex. The contributions    built from higher-pt correlators are automatically $N$-suppressed and thus negligible. }}
\be \langle{\cal O}{\cal O}\rangle_{J}=\frac{1}{\langle{\cal O}{\cal O}\rangle^{-1}+i\xi} \,.\ee
Like in the weakly coupled case, this can be  derived from the partition function, which produces a Born series representing the two-point CFT correlator dressed by insertions of $-iJ$.
If the ${\cal O}$ operator satisfies \eqref{eq:bound}, the dressed correlator takes the form 
\be \langle{\cal O}{\cal O}\rangle_{J}=-\frac{i}{\xi}+\frac{1}{\xi^2}\langle{\cal O}{\cal O}\rangle^{-1} +O(\xi^{-3}) \,\ee
in the IR. The first term is a mere contact term. 
The second term features the inverse 2-point correlator, that turns out to be proportional to the 2-point correlator of an operator ${\tilde {\cal O}}$ with dimension $\tilde \Delta=d-\Delta$ with $\frac{d}{2}<\tilde \Delta<\frac{d}{2}+1$. 
One says that the deformations induces a RG flow from a UV CFT  with an operator of dimension $\Delta$ to an IR CFT with an operator of dimension $d-\Delta$. See \cite{Porrati:2016lzr} and references therein, and the seminal papers \cite{ Brezin:1972se,Wilson:1973jj}.

Let us now model  imperfect defects and boundaries in CFT via a \textit{localized} relevant double-trace deformation. {Like in the weakly coupled case, the two-point correlator can be expressed as a  Born series.}
To express it rigorously in position space, we introduce the convolution product  $\cB$ as $f\cB g(x_1,x_2) = \int d^d xf(x_1,x) g(x,x_2)$ and  introduce the inverse
\be
A\cB A^{-1}(x) = \delta^d(x) \,.
\ee
We also introduce $\Sigma(x,x')=-i J(x) \delta^d(x-x')$. Using this notation we can write the propagator entirely using convolutions. 
The exact resummed Born series is expressed as 
\begin{eqnarray}
\langle{\cal O}(x_1){\cal O}(x_2)\rangle_{J} 
&= \sum_{r=0}^\infty  \langle{\cal O}{\cal O} \rangle \left[\star \,\xi \Sigma \star \langle{\cal O}{\cal O} \rangle  \right]^{r} (x_{12})
\\ &=\left[ \langle{\cal O}{\cal O}\rangle^{-1}-\xi \Sigma 
\right]^{-1}(x_{12})\,. \label{eq:G_gen_sum}
\end{eqnarray}
If ${\cal O}^2$ is relevant, then in the infrared the $\xi$ term must dominate at any point of the $J$ support. 
We thus obtain that,  for any $x_{1}$ or $x_2$ in $J$, 
\be \langle{\cal O}(x_1){\cal O}(x_2)\rangle_{J} = \frac{1}{\xi}\delta^d(x_{12}) +\frac{1}{\xi^2}
\langle{\cal O}(x_1){\cal O}(x_2)\rangle^{-1}  +O(\xi^{-3})  \,. 
\ee
We can see that the deformed CFT 2-point correlator tends not to  propagate inside $J$ in the infrared regime. { Asymptotically in the IR, when  $\xi\to\infty$, we obtain that  $\langle{\cal O}(x_1){\cal O}(x_2)\rangle_{J}\to 0$  anywhere on $J$ and its boundary. Therefore the 2-point correlator satisfies a Dirichlet condition on the boundary of $J$  in the IR. } Such a behavior appropriately models an imperfect defect/boundary for a  CFT.

\subsection{The Double-Trace Membrane}
\label{se:membrane}

A simple extended double-trace defect is the one whose  support is a codimension-one plane. We refer to it as a  membrane.
The support of the membrane is defined as\,\footnote{{From now on we include the coupling constants $\xi$ in $J$.}}
\be J(\x)=\xi\delta(z-z_0) \label{eq:J_membrane}\,. \ee
To compute the dressed propagator, 
 one uses the position-momentum space 2-point correlator  Eq.\,\eqref{eq:2pt_mixed}.  Dressing the propagator with a membrane necessarily involves evaluating $\llangle  {\cal O}(p,z_1)  {\cal O}(-p,z_2) \rrangle $ at $z_{12}=0$. Let us investigate its behavior for small $z_{12}$ {at fixed $p$. In this limit 
 the Bessel function has a small argument expansion.  }
 We find
{\be
\llangle  {\cal O}(p,z_1)  {\cal O}(-p,z_2) \rrangle_d =  \left(
\llangle  {\cal O}(p)  {\cal O}(-p) \rrangle_{ d-1} + \frac{c}{z^{2\Delta-d+1}_{12}} \right)\left[1+ O\left((pz_{12})^2\right)\right] 
\label{eq:2point_smallz12}
\ee }
with
\be
c= -i \frac{\Gamma(\Delta+\frac{1-d}{2})}{\Gamma(\Delta)}\,. 
\ee

{ The two terms shown in Eq.\,\eqref{eq:2point_smallz12} are the leading non-analytical and analytical ones.  These two terms   correspond respectively to the regions of small and large $p_z$ momentum covered by the corresponding Fourier integral. }
The $\llangle  {\cal O}(p)  {\cal O}(-p) \rrangle_{ d-1} $ correlator, which is independent of $z_{12}$, corresponds exactly to the 2-point correlator of an operator of dimension $\Delta$ in $d-1$ dimensions. One could equivalently obtain it by averaging over $z_{12}$ in the original position space correlator. 

The $\frac{c}{z^{2\Delta-d+1}_{12}}$ term corresponds to large $p_z$ momentum. One could equivalently obtain it by averaging the transverse coordinates in the original position space correlator. 
One can see  that this term diverges when $z_{12}\to0$ if $\Delta> \frac{d-1}{2}$. {This divergence might need to be treated via renormalization of the defect. 
This would deserve a separate treatment that is beyond the scope of this note. }Therefore, in the presence of a membrane, we restrict $\Delta$ as 
\be
\frac{d-2}{2}\leq\Delta < \frac{d-1}{2}\,. \label{eq:bound2}
\ee

We denote the 2-point function in the presence of the defect $J$ as \be \llangle  {\cal O}(p,z_1)  {\cal O}(-p,z_2) \rrangle_J \equiv i G_J(p;z_1,z_2)\,.\ee 
In the case of the membrane \eqref{eq:J_membrane}, we obtain
\be
G_J(p;z_1,z_2) = G(p;z_1,z_2)+
G(p;z_1,z_0)\frac{\xi}{1 -\xi G_0(p)}G(p;z_0,z_2)\,. \label{eq:2point_membrane}
\ee
where $G_0(p)=G(p;z_0,z_0) = \llangle  {\cal O}(p)  {\cal O}(-p) \rrangle_{ d-1}$ corresponds to the 2-point function in $d-1$ space defined in \eqref{eq:2point_smallz12}.  Explicitly,
\be
G_0(p) =  -  \frac{\pi^{\frac{d-1}{2}} \Gamma(\frac{d-1}{2}-\Delta)}{\Gamma(\Delta)}\left(\frac{4}{-p^2}\right)^{\frac{d-1}{2}-\Delta}
\label{eq:G0}\,.
\ee
If the double-trace operator is relevant, $ G_0(p)$ grows when $p$ decreases. In the limit for which $\xi G_0(p)\gg 1$, we have therefore 
\be
G_J(p;z_1,z_2) \xrightarrow[{\rm small}~ p ]{} G(p;z_1,z_2)-
G(p;z_1,z_0)G_0^{-1}(p) G(p;z_0,z_2) \label{eq:Dir_limit_membrane}
\ee
which satisfies Dirichlet boundary condition on the membrane. 

The membrane defect  can serve as  an approximation for a plate-shaped defect of finite width. The approximation appears in the IR regime, when 
the plate width is smaller than all other distance scales of the problem such that, by dimensional analysis, the correlator must see the plate 
 approximately as a membrane.

\subsection{AdS/CFT Motivation}

Another motivation for implementing relevant double-trace deformations as defects and boundaries comes from the
AdS/CFT correspondence. See \cite{Aharony:1999ti,Zaffaroni:2000vh,Nastase:2007kj,Kap:lecture} 
for some AdS/CFT reviews and lecture notes. 

Let us consider the $d+1$-dimensional Poincaré patch with a boundary at $y=y_0$, $ds^2=\frac{L^2}{y^2}(dx^\mu  dx_\mu+d y^2)$, $y\geq y_0$.  Consider a scalar field in the bulk of AdS with mass $m_\Phi^2=\Delta(\Delta-d)L^2$. 
For any $\Delta>\frac{d-2}{2}$, the brane-to-brane propagator of $\Phi$ behaves as the one of a $d$-dimensional free field $\phi$ mixing  with the 2-point function of a CFT operator of dimension $\Delta$ via an operator $\phi {\cal O}$. The same is true for higher point correlators. This is sometimes referred to as the $\Delta_+$ branch of the correspondence.

When $\Delta<\frac{d}{2}$, a second possibility appears: the brane-to-brane correlators can be directly identified  as the CFT correlators of an operator with dimension $d-\Delta$. See \cite{Klebanov:1999tb,Mueck:2002gm} and e.g. \cite{Porrati:2016lzr,Giombi:2018vtc, Geng:2023ynk} for more recent works. We refer to this identification as the $\Delta_-$ branch of the correspondence. Here we write the general statement of the $\Delta_-$ branch as \be
\int {\cal D}\varphi_{\rm CFT} e^{iS_{\rm CFT}+iS_0[{\cal O},J]}\equiv \int {\cal D}\Phi_0 e^{ iS_0[\Phi_0,J]}
\int_{\Phi_0} {\cal D}\Phi e^{iS_{\rm AdS}[\Phi]} \label{eq:AdSCFT_minus}
\ee
where $\Phi_0$ denotes the value of the fields on the boundary, here $\Phi_0=\Phi|_{z=z_0}$.\,\footnote{ The $S_0$ action can contain a linear source term $S_0[X,\bar J]=\int d^{d}x X \bar J$, that can be used to define the correlators on both sides upon functional derivative in $\bar J$.} 

In our model of defect CFT, the general double-trace deformation Eq.\,\eqref{eq:DTdef} corresponds to setting the $S_0$ action to 
\be
S_0[X,J]\equiv -\frac{\xi}{2}\int d^dx X^2 J\,.
\ee
Using Eq.\,\eqref{eq:AdSCFT_minus}, we see that this corresponds to a boundary-localized mass term for $\Phi$ on the AdS side. Therefore, the double-trace deformation on the CFT side is encoded as a deformation of the boundary condition of $\Phi$ on the AdS side. The  double-trace defect of the CFT is realized as a boundary mass term with a support that is localized along the boundary volume. { In short, the defect is on the boundary:}
\be
	\includegraphics[width=0.6\linewidth,trim={5cm 4.7cm 6cm 4.3cm},clip]{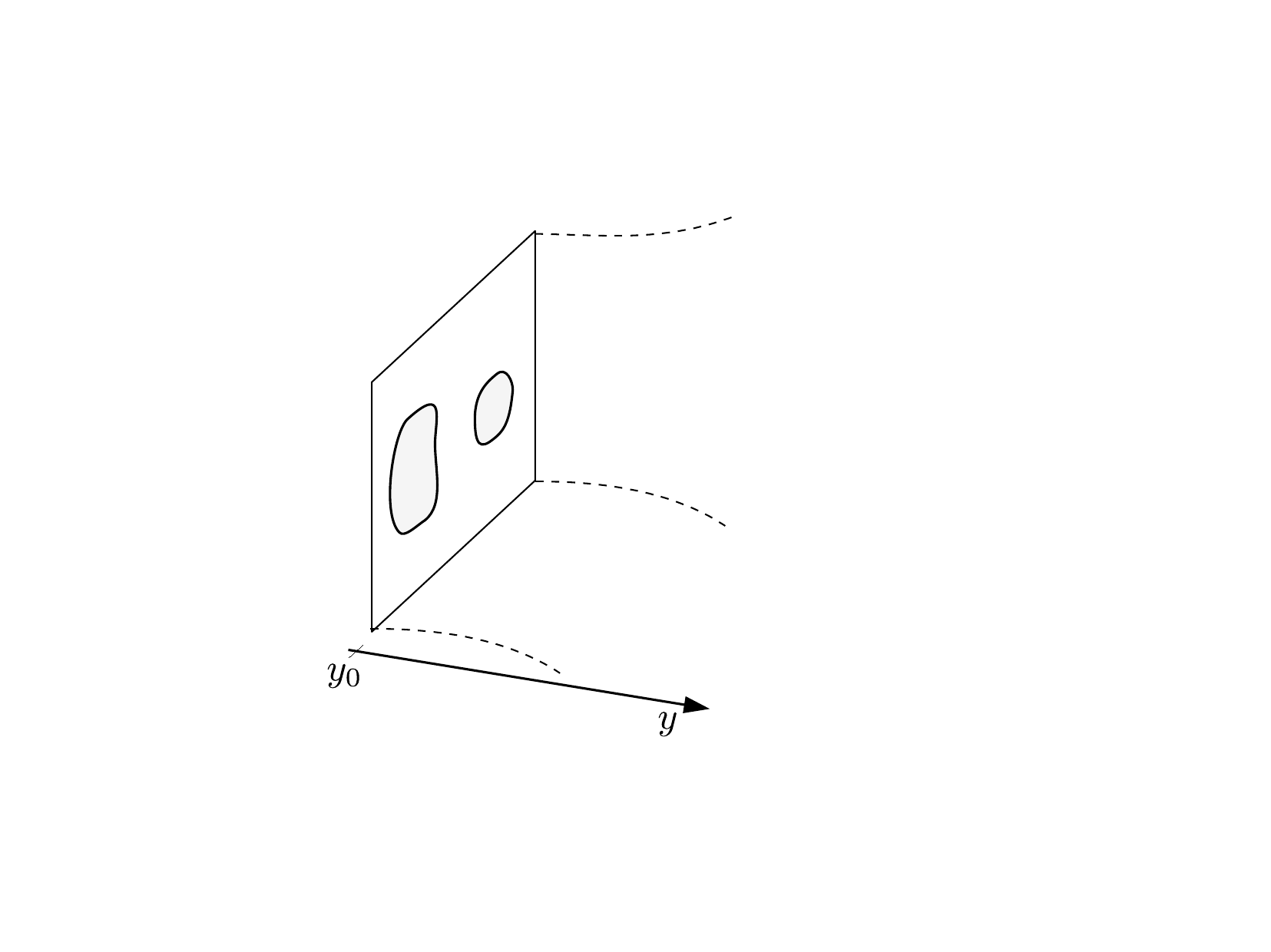}
\nn
\ee

The domain for which the $\Delta_-$ correspondence applies is precisely the range given in Eq.\,\eqref{eq:bound}. Thus our model of defect CFT can always be realized holographically  from the AdS viewpoint:  The   double-trace deformation  defined  via AdS is automatically relevant.

At the level of the vacuum energies we  have the identification
\be
E_{\rm CFT}[J]= E_{\rm AdS}[J] \label{eq:EJ_AdSCFT}
\ee
with \be
 E_{\rm AdS}[J] = iT\log Z_{\rm AdS}[J]\,,\quad\quad 
 Z_{\rm AdS}[J]= \int {\cal D}\Phi_0 e^{i \int_\partial d^dx \frac{\xi}{2}J(\x)\Phi_0^2}
\int_{\Phi_0} {\cal D}\Phi e^{iS_{\rm AdS}[\Phi]}\,. 
\ee
When the correspondence Eq.\,\eqref{eq:EJ_AdSCFT} holds,  
applying the functional formalism of section \ref{se:func} to $E_{\rm CFT}$ means on the AdS side that we deform the support of the boundary-localized mass term. In other words,  the boundary condition for the bulk fields gets deformed.
The phenomenon of the 2-point correlator being repelled from the defect in the IR is understood on the AdS side as the bulk field being repelled from the boundary due to the  mass term. 
For $\xi \to \infty$, the AdS propagator vanishes  on the boundary of the defect localized on the AdS boundary. 

We do not use further the AdS picture in the following.

\section{A CFT between Two Membranes}

\label{se:membranes}

We explore further the properties of the 2-point CFT correlators in the presence of two double-trace membranes. 
The full defect is given by \be
J(\x) = J_a(\x)+ J_b(\x) = \frac{\xi_a}{2}\delta(z-z_a)+ \frac{\xi_b}{2}\delta(z-z_b)\,.
\ee
We define $|z_b-z_a|=L$.

A convenient way to obtain the two-point function is by dressing it successively with the two membranes $J_a$ and $J_b$. We obtain 
\be
G_a(p;z_1,z_2)= G(p;z_1,z_2)+
G(p;z_1,z_a)\frac{\xi_a}{1-\xi_a G(p;z_a,z_a)}G_0(p;z_a,z_2)\,,
\label{eq:dressed_one_membranes}
\ee
\begin{eqnarray}
&& G_{a,b}(p;z_1,z_2)=  G_a(p;z_1,z_2)+
G_a(p;z_1,z_b)\frac{\xi_b}{1-\xi_b G_a(p;z_b,z_b)}G_a(p;z_b,z_2)\, \label{eq:dressed_two_membranes} \\
&& = G_{12}+ \frac{
\xi_b G_{1b}\left(\xi_a G_{ab} G_{a2}  +(1-\xi_a G_0) G_{b2} \right)  + \xi_a G_{1a}(\xi_bG_{ab}G_{b2}+ (1 -\xi_b G_0) G_{a2}   )
}{(\xi_a G_0-1)(\xi_b G_0-1)-\xi_a\xi_b G^2_{ab}} \nn 
\end{eqnarray}
In the last line we introduced the notation $G(p;z_i,z_j)\equiv G_{ij}$.

\paragraph{Dirichlet limit.} 
To understand the behavior of this 2-point function, we take the $\xi_{a,b}\to\infty$. At finite $\xi_{a,b}$, this corresponds to the  asymptotic limit associated to the infrared regime. 
In  this limit 
the CFT gets literally confined inside the $[0,L]$ interval. 
The 2-point correlator becomes 
\be
 G_{\rm D}(p;z_1,z_2)= 
 G_{12}+ \frac{
 G_{1b} G_{ab} G_{a2}  - G_{1b} G_0 G_{b2}   +  G_{1a} G_{ab}G_{b2}  - G_{1a} G_0 G_{a2}   
}{ G^2_0- G^2_{ab}}\,. \label{eq:G_Dir}
\ee

\paragraph{Poles.} 

We stay in the Dirichlet limit for simplicity. 
Due to the denominator in Eq.\,\eqref{eq:G_Dir}, it turns out that 
$G_D$ features a series of  poles in the complex plane of $p$ determined by the condition
\be
G(p;z_a,z_b)=\pm G_0(p)\,. \label{eq:poles1}
\ee

Explicitly, the poles in  $p$ are determined by solving 
\be
\frac{1}{\Gamma(\alpha)}\left ( \sqrt{-p^2}L\right)^\alpha K_\alpha (\sqrt{-p^2}L)=\pm 1, \ \ \ \alpha= \frac{d-1}{2}-\Delta
\label{eq:poles2}
\ee
with $-\frac{1}{2}<\alpha < \frac{1}{2}$.
We denote the complex values of $p$ solving Eq.\,\eqref{eq:poles2} by $m_n^\pm$.  
There is no massless pole ($p=0$) nor light pole ($p\ll \frac{1}{L}$), because the asymptotic behavior at small $p$ is $G_{ab}\to G_0$ (see \eqref{eq:2point_smallz12}),  in which case  Eq.\,\eqref{eq:poles1} is either trivial or impossible to satisfy.

\paragraph{Residues.} 
 The residues associated to the $p=m_n^\pm$ poles  take a simple factorized form, $(G_{a1}\mp G_{b1})(G_{a2}\mp G_{b2} )$, 
\be
G(p;z_1,z_2) \overset{p\sim m_n^\pm}{\approx} -\frac{1}{2}\frac{f^\pm_n(z_1)f^\pm_n(z_2)}{G_{ab}\mp G_{0}}\,,\quad\quad f_n(z)\equiv 
G(m_n^\pm,z,z_a)\mp G(m_n^\pm,z,z_b) \,. \label{eq:G_conf}
\ee
This factorized form is reminiscent of weakly coupled QFT on an interval, which develops a sequence of discrete modes. In the weakly coupled case,  the poles lie on the real line up to corrections due to the interactions. Eq.\,\eqref{eq:G_conf} then corresponds to the Kállen-Lehmann representation of the propagator confined in the $[0,L]$ interval. 
Here, we see that the factorized structure remains true even if the poles lie anywhere in the complex plane.

\paragraph{Free limit.} 
In the case of the free field in $d=4$, we have $\Delta=1$. The two-point correlator  becomes
\be i G^{\rm free}(p;z_1,z_2) = -i4\pi^2 a \frac{e^{-\sqrt{-p^2}|z_1-z_2|}}{2 \sqrt{-p^2}}\,
\ee 
where for a canonically normalized field  $a=\frac{1}{4\pi^2}$. 
In this case the poles determined by Eq.\,\eqref{eq:poles2} are real, with $m^{\rm free}_n=n\pi/L$, $n\in \mathbb{N}^\star$.  
The propagator dressed by the two membranes takes the form 
\be
i G^{\rm free}(p;z_1,z_2) =  i\frac{
\sinh\left(\sqrt{-p^2}(z_a-z_<)\right)
\sinh\left(\sqrt{-p^2}(z_>-z_b)\right)
}{\sqrt{-p^2} \sinh\left(\sqrt{-p^2}(z_b-z_a)\right)}\,
\ee
where we assumed $z_a<z_b$ and defined $z_{<(>)}=\min (\max) (z,z')$.   
This matches the result obtained by solving the free field  equation of motion on the interval with Dirichlet boundary conditions on the membranes (see e.g. appendix of \cite{Fichet:2019owx}).

 \paragraph{Resonances.} 
Slightly away from the free field case, for $\Delta-\frac{d-2}{2}\ll 1$, it turns out that the set of poles of the CFT behaves as a tower of narrow resonances, at values $p=m_n\equiv m^{\rm free}_n-i\Gamma_n/2$ with $\Gamma_n\ll m^{\rm free}_n$. Expanding the relation   \eqref{eq:poles2} we find that the resonances feature a common decay rate $\Gamma_n$
\be
\Gamma_n\approx\frac{(\Delta-\frac{d-2}{2})\pi}{L}\,. 
\label{eq:widths}
\ee
Details of the computation are given in Appendix \ref{app:widths}. 

We  obtain thus a notion of unstable particle states directly from a CFT. 
Since the CFT has internal degrees of freedom, one may think of these resonances as collective excitations. The fact that the resonances decay reflects the fact that, for $\Delta>\frac{d-2}{2}$, the theory is interacting. However the decay width is independent of the underlying physics of the CFT, it is controlled by the dimension of the double-trace operator that causes the CFT confinement on the interval.

\section{CFT Casimir forces between Defects and Boundaries }

\label{se:Casimir}

We compute the quantum forces induced by the CFT between localized double-trace operators with pointlike and planar supports. 
The planar geometry includes the case of a flat boundary (e.g. $z>0$), of a membrane, and also the case of a plate of any width. 
We consider two disjoint defects, described by
\be
S=S_{\rm CFT}-\frac{1}{2}\int d^dx {\cal O}^2(x) J(\x)\,\quad\quad\quad  J(\x)=\xi_a J_a(\x) + \xi_b J_b(\x)\,. \label{eq:twodefects}
\ee
{The $\xi_{a,b}$ parameters have mass dimension $[\xi_{a,b}]=d-[J_{a,b}]-2\Delta$}. 

We consider a rigid deformation of $J$ such that $J_b$ gets shifted along a constant $\Lv$ while $J_a$ remains identical,
\be
J_{a,\lambda+d\lambda}(\x)=J_{a,\lambda}(\x )\,,\quad\quad J_{b,\lambda+d\lambda}(\x)=J_{b,\lambda}(\x-\Lv d\lambda )\,, \label{eq:deformation_gen}
\ee
{i.e.}
 \be
	\includegraphics[width=0.7\linewidth,trim={5cm 8.cm 3cm 7cm},clip]{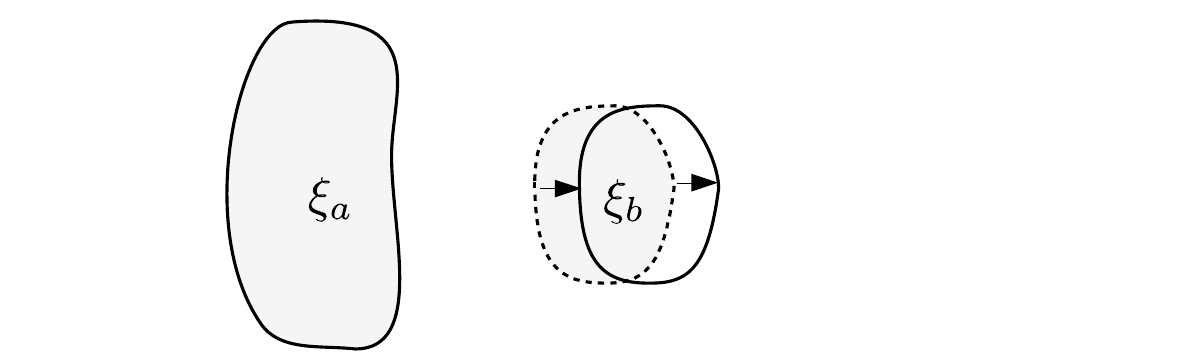}
 \nn
 \ee
The quantum work is then expressed as 
\be
W=  -\frac{\xi_b}{2} \int d^{d-1} \x  \langle {\cal O}(x) {\cal O}(x) \rangle_J  \partial_\lambda J_{b,\lambda}(\x) \label{eq:Work_Casimir}
\ee
This is the formula we apply throughout this section.

The  CFT propagator in the presence of $J$ can always be written in the form of a Born series as described in Eq.\,\eqref{eq:G_gen_sum}. 
Evaluating the expression in a closed form for e.g. a plate, is more challenging. 
Here we limit ourselves to computing analytical results for the force between $J_a$ and $J_b$ in two limiting cases: the asymptotic Casimir-Polder and Casimir regimes.

\subsection{CFT  Casimir-Polder Forces}

\label{se:CasimirPolder}

 In the UV regime, i.e. in the limit of short separation, the effect of the $J$ insertion in the Born series tends to be small. In this limit the first terms of the series dominate. It turns out that the leading contribution to the quantum work is  \cite{Brax:2022wrt}
\begin{eqnarray}
W & =&  -i \frac{\xi_a\xi_b}{2}  \int d^{d-1}\x d^dx' \langle {\cal O}(x') {\cal O}(x) \rangle  J_a(\x)\langle {\cal O}(x) {\cal O}(x') \rangle \Lv\cdot \bpartial J_b(\x') +O(\xi^3)\,.  
\end{eqnarray}
Upon integration by part, we recognize the structure of a potential,  $W=-\Lv \cdot \bpartial V_{ab}$ with 
\be
V_{ab}= - i \frac{\xi_a\xi_b}{2} \int d^{d-1}\x d^{d-1}\x' J_a(\x)J_b(\x') \int dt \,\langle {\cal O}(0,\x) {\cal O}(t,\x') \rangle ^2\,.
\ee
The $V_{ab}$ potential  has a Casimir-Polder-like structure,  it is a loop made of two CFT correlators that connects the two defects. Hence we refer to this limit as ``Casimir-Polder''.

\subsubsection{Point-Point geometry}

We first consider  two defects that are pointlike, 
 \be
 J_a({\bf x }) =  \delta^{d-1} ({\bf x})\,,\quad\quad 
 J_b({\bf x }) =  \delta^{d-1} ({\bf x}-{\bf r})\,.
 \ee
The potential becomes 
\be
V(r)= - i \frac{\xi_a\xi_b}{2}  \int dt \,\langle {\cal O}(0) {\cal O}(t,r) \rangle ^2\,
\label{eq:V_CP_pointpoint}
\ee
with $r=|{\bm r}|$. 

We compute Eq.\,\eqref{eq:V_CP_pointpoint} by going to  full momentum space. The momentum space correlator is Eq.\,\eqref{eq:2pt_momentum}. 
In momentum space the potential is given by 
\be
V(p) = i \frac{\xi_a\xi_b}{2} 
\frac{\pi^{d} \Gamma^2(d/2-\Delta)}{\Gamma^2(\Delta)} 
\int \frac{d^{d} k}{(2\pi)^{d}} 
\left(\frac{4}{-k^2}\right)^{d/2-\Delta} \left(\frac{4}{-(k+p)^2}\right)^{d/2-\Delta}
\ee
where $p_0=0$. We rotate the integral to Euclidian space with Euclidian momentum $q^M$ satisfying $q^2=-k^2$, and go to spherical coordinates, 
\be
V(p) = - \frac{\xi_a\xi_b}{2}  
\frac{\pi^{d} \Gamma^2(d/2-\Delta)}{\Gamma^2(\Delta)} 
\int \frac{d^{d} q}{(2\pi)^{d}} 
\left(\frac{4}{q^2}\right)^{d/2-\Delta} \left(\frac{4}{(q+p)^2}\right)^{d/2-\Delta}
\ee

We need to evaluate
\begin{align}
    \int\frac{d^dq}{(2\pi)^d} \left(\left(p+q\right)^2\right)^a \left(q^2\right)^b \label{eq:int1}
\end{align}
for some $a,b$. We apply the identity
\begin{align}\label{eq:FIidentity}
\left(\left(p+q\right)^2\right)^a \left(q^2\right)^b=\int_0^1 dx\frac{(x\left(p+q\right)^2+(1-x)q^2)^{a+b}}{x^{a+1}(1-x)^{b+1}}\frac{\Gamma(-a-b)}{\Gamma(-a)\Gamma(-b)}\,.
\end{align}
The integral on the right-hand side converges for  ${\rm Re}(a),{\rm Re} (b)<0$.
However, provided the final result of the calculation is analytic in $a,b$, the result can be extended by analytical continuation such that restrictions on $a,b$ are ultimately lifted.
Shifting the loop momentum $l\equiv q+px$, we obtain
\begin{align}
 \eqref{eq:int1} =   \int_0^1 dx\int\frac{d^dl}{(2\pi)^d} \frac{(l^2+x(1-x)p^2)^{a+b}}{x^{a+1}(1-x)^{b+1}}\frac{\Gamma(-a-b)}{\Gamma(-a)\Gamma(-b)}.
\end{align}
We evaluate the loop integral with
\begin{align}
\int\frac{d^dl}{(2\pi)^d}\left(l^2+\Delta\right)^c=\frac{\Gamma\left(-c-\frac{d}{2}\right)}{\Gamma(-c)} \frac{\Delta^{c+\frac{d}{2}}}{(4\pi)^{\frac{d}{2}}} \,.\label{eq:loopintegral}
\end{align}

Again, the loop integrals are performed in the domain of $(c,d)$ where the integral on the left-hand-side converges. The functions on the right-hand-side are analytic in $c$ anywhere away from integral values of $c$, hence the final result will be ultimately analytically continued in $c$. 
For certain  values of $\Delta$  at even $d$,  a physical divergence appear which requires renormalization. However such divergences are irrelevant for our study because it is ultimately only the branch cut of $V(p)$ that contributes to the spatial potential, see \cite{Brax:2017xho}.  Hence  no divergence appears in the position space propagators when $\Delta$ is set to integer values.

Putting Eqs.\,\eqref{eq:FIidentity} and \eqref{eq:loopintegral} together yields
\begin{align}
\label{eq:Pi_beta}
    \eqref{eq:int1} =   \frac{1}{(4\pi)^{\frac{d}{2}}}\left(p^2\right)^{a+b+\frac{d}{2}}\frac{\Gamma\left(-a-b-\frac{d}{2}\right)}{\Gamma(-a)\Gamma(-b)}\int_0^1 dxx^{b+\frac{d}{2}-1}(1-x)^{a+\frac{d}{2}-1}.
\end{align}
We identify the remaining integral as being the integral representation of the Beta function. Evaluating the integral, we obtain
\begin{align}\label{eq:eq411}
   \eqref{eq:int1} =   \frac{1}{(4\pi)^{d/2}}\left(p^2\right)^{a+b+d/2}\frac{\Gamma(-a-b-d/2)}{\Gamma(-a)\Gamma(-b)}\frac{\Gamma(a+d/2)\Gamma(b+d/2)}{\Gamma(a+b+d)}.
\end{align}

The potential in momentum space is thus 
\be
V(p) = - \frac{\xi_a\xi_b}{2}  
\frac{\pi^{d} \Gamma^2(d/2-\Delta)}{\Gamma^2(\Delta)} 
4^{d-2\Delta}
\frac{1}{(4\pi)^{d/2}}\left(p^2\right)^{2\Delta-d/2}\frac{\Gamma(-2\Delta+d/2)}{\Gamma(\frac{d}{2}-\Delta)\Gamma(\frac{d}{2}-\Delta)}\frac{\Gamma^2(\Delta)}{\Gamma(2\Delta)}.
\ee
Simplifying, 
\be
V(p) = - \frac{\xi_a\xi_b}{2} 
\left(\frac{p^2}{4}\right)^{2\Delta-d/2}\frac{\pi^{d/2}\,\Gamma(-2\Delta+d/2)}{\Gamma(2\Delta)}\,. \label{eq:Vp_pointpoint}
\ee

We  can  recognize that Eq.\,\eqref{eq:Vp_pointpoint} is proportional to the momentum space 2-point correlator of the double-trace operator ${\cal O}^2$ with $p_0=0$.  That is, due to the properties of the CFT, the loop of ${\cal O}$ can be understood as a tree exchange of ${\cal O}^2$.\,\footnote{
 Particle physics models involving such processes   have been considered in \cite{Brax:2019koq,Costantino:2019ixl,Chaffey:2021tmj}. } The overall coefficient is nontrivial, however, our loop calculation is required to determine it. This phenomenon occurs only in the  Casimir-Polder regime.

One may notice that the numerator diverges  if $\Delta\to \frac{d}{4}$, which is allowed when $d\leq 4$ since $\Delta\geq \frac{d-2}{2}$. However, the expression for the potential in position space computed below is automatically finite even in the case $\Delta\to \frac{d}{4}$. This is because this is a quantity computed at separated points.  Keeping a general, non-integer, dimension $\Delta$ throughout the calculation plays the same role as dimensional regularization weakly coupled QFT.  
Finally we can go back to  position space with a $d-1$ Fourier transform,
$V(r)=\int \frac{d^{d-1}p}{(2\pi)^{d-1}} e^{i p r} V(p) $.
We obtain the final result for the CFT Casimir-Polder force between two pointlike double-trace deformations,
\be
V(r) = - \sqrt{\pi} \frac{\xi_a\xi_b}{2}
\frac{\Gamma(2\Delta -\frac{1}{2})}{\Gamma(2\Delta)}
\frac{1}{r^{4\Delta-1}}
\ee
As a cross-check, taking $\Delta=1$ and using  the $a=\frac{1}{4\pi}$ normalization for each correlator, we recover exactly the Casimir-Polder force from the exchange of 4D  free massless scalars, $V(r)= -\frac{\xi^2}{64\pi^3r^3}$. {Notice that  $[\xi_{a,b}]=1-2\Delta$ thus $[V]=1$.}

\subsubsection{Point-Plate}
\label{se:CP_pointplate}

We  calculate the Casimir-Polder potential between a point particle and an infinite plate located at $z<0$. 
In terms of the support functions, this is described by
 \be
 J_a({\bf x }) =  \Theta(-z)\,,\quad\quad 
 J_b({\bf x }) =  \delta^{d-2} ({\bf x}_\parallel)\delta(z-\ell)\,.
 \label{eq:source_pointplate}
 \ee
We assume that the deformation moves $J_b$ along the $z$ direction, i.e. $\Lv=({\bf 0},1)$. The $x_\parallel$ are the coordinates parallel to the plate. 

The  CFT force between the point and the membrane can be easily obtained by integrating the point-point Casimir-Polder potential over $J_a$. This simple approach is valid only in the Casimir-Polder limit. 
The $\xi_{a,b}$ are defined such that the parametrization  Eq.\,\eqref{eq:twodefects} holds, now with the defect \eqref{eq:source_pointplate}. 
$\xi_a$ is related to the pointlike source coupling by 
$\xi_a=n \xi^{\rm point}_a$ where $n$ is the number density of $J_a$.

The Casimir-Polder force is given by 
\be 
V(\ell)= n \int_{-\infty}^0 dz \int d^2 x_{\parallel} \int \frac{d^3p}{(2\pi)^3} e^{ip_z(\ell-z)} e^{ip_\parallel.x_{\parallel}} V(p)\,. \label{eq:Vlint1}
\ee
The $p_\parallel$ are the momentum component along the plate. The integral reduces to
\be 
V(\ell)= - \frac{\pi^{d/2}\,\xi_a\xi_b}{2}  \frac{\Gamma(-2\Delta+d/2)}{\Gamma(2\Delta)}
   \int_{-\infty}^0 dz \int \frac{dp_z}{2\pi}  e^{ip_z(\ell-z)} 
\left(\frac{p^2}{4}\right)^{2\Delta-d/2}
\ee
The momentum integral can be performed and gives
\be
\int \frac{dp_z}{2\pi}  e^{ip_z(\ell-z)} 
\left(\frac{p^2}{4}\right)^{2\Delta-d/2}= \frac{\Gamma(2\Delta+\frac{1-d}{2})}{\sqrt{\pi}\Gamma(\frac{d}{2}-2\Delta)}\frac{1}{(\ell -z)^{4\Delta+1-d}}
\ee
The integral over $z$ converges provided $\Delta > d/4$.
When computing the force further below, the divergence matters only when  for a free field in $d=3$. 
In the convergent case we have
\be 
V(\ell)= - \frac{\pi^{(d-1)/2}}{2(4\Delta -d)}  \frac{\Gamma(2\Delta+\frac{1-d}{2})}{\Gamma(2\Delta)}
  \frac{\xi_a\xi_b}{\ell^{4\Delta -d}}
\ee
where $d>4$. 
The force is then given by $F=-\frac{\partial V}{\partial \ell}$, which gives
\be 
F(\ell)= - \frac{\pi^{(d-1)/2}\,\Gamma(2\Delta+\frac{1-d}{2})}{2\,\Gamma(2\Delta)}
  \frac{\xi_a\xi_b}{\ell^{4\Delta -d+1}}\,. \label{eq:F_CP_pointplate}
\ee
For  a free field in $d=4$ we obtain 
\be
F(\ell)= -\frac{\pi^2}{2} 
  \frac{\xi_a\xi_b}{\ell} \,. 
\ee
This correctly reproduces the $\propto \frac{1}{\ell}$ scalar Casimir-Polder force derived in \cite{Brax:2022wrt}, $- \frac{\xi_a\xi_b}{32\pi^2\,\ell}$,  once one takes into account the canonical normalization of the free fields, which introduces the factor $a^2=(\frac{1}{4\pi^2})^2$.\,\footnote{A factor of $\frac{1}{2}$ is missing in Eq.\,(6.29) of \cite{Brax:2022wrt}.}

The case of the free field in $d=3$ necessitates to assume that the plane has finite width $L$. We obtain
\be
F(\ell)= -\frac{\pi \,\Gamma(2\Delta-1)}{2\,\Gamma(2\Delta)}
  \xi_a\xi_b \log\left(1+\frac{L}{\ell}\right)\,.
\ee

\subsubsection{Plate-Plate CFT  Casimir-Polder}

We similarly compute the Casimir-Polder {pressure} between two infinite plates. This is described by 
 \be
 J_a({\bf x }) =  \Theta(-z)\,,\quad\quad 
 J_b({\bf x }) =  \Theta(z-\ell)\,.
 \label{eq:source_plateplate}
 \ee
 We assume that the deformation moves $J_b$ along the $z$ direction, i.e. $\Lv=({\bf 0},1)$.\,\footnote{
In general one should require that the plates end far away, i.e. are not formally infinite,  in order for the deformation flow to be divergence-free \cite{Brax:2022wrt}.  While it is necessary in general, this detail does not affect the present Casimir-Polder calculation. 
 }

 The $\xi_{a,b}$ are defined such that the parametrization  Eq.\,\eqref{eq:twodefects} holds, now in the presence of the defect \eqref{eq:source_pointplate}. 
$\xi_{a,b}$ is related to the pointlike source coupling by 
$\xi_{a,b}=n_{a,b} \xi^{\rm point}_{a,b}$ with $n_{a,b}$ the number density of $J_{a,b}$. 
 Similarly to the point-plate case, we integrate the point-point potential over the two defects, with e.g.
\be 
F_{\rm plate-plate}(\ell)= -n_b S_{d-2} \int_{\ell}^\infty dz F_{\rm point-plate}(z) \label{eq:plate-plate_integral}
\ee
where  $S_{d-2}=\int d^{d-2}\x_\parallel$ is the  volume integral in the directions parallel to the plate. 
As long as $\Delta > d/4$, the integral is IR convergent and gives
\be 
\frac{F(\ell)}{S_{d-2}}=-
\frac{\pi^{(d-1)/2}}{2(4\Delta -d)} \frac{\Gamma(2\Delta+\frac{1-d}{2})}{\Gamma(2\Delta)}
  \frac{\xi_a\xi_b}{\ell^{4\Delta -d}}\,.
\ee

The case of a free field in $d=4$ is logarithmically divergent. This is a physical divergence that signals that we should consider finite plates instead of approximating them as infinite. It is sufficient to  assume that one of the plates, here the second plate integrated in Eq.\,\eqref{eq:plate-plate_integral}, has finite width $L$. We find
\be 
\frac{F(\ell)}{S_1}=
-\frac{\pi^{2}}{2}\xi_a\xi_b  \ln \left( 1+\frac{L}{\ell}\right)\,.
\ee
The infrared divergent behavior also appears in the result of \cite{Brax:2022wrt}, 
in the case where the free field is massless. There is no IR divergence if the free field is massive.

\subsection{CFT  Casimir Forces}
\label{se:CasimirCasimir}

We compute forces beyond the Casimir-Polder approximation. Our focus is on membranes. 
Computing analytical results for plates of finite widths is more challenging. However, in the IR regime for which the plate width is smaller than other distance scales of the problem, we expect the results to reproduce the one obtained with membranes.

Since the chosen defects feature membranes, we restrict $\Delta$ to the interval given in Eq.\,\eqref{eq:bound2}. The case $\Delta\geq \frac{d-1}{2}$ would deserve a separate analysis.

\subsubsection{Point-Membrane CFT Casimir}

We first compute the force between a membrane at $z=0$ and a point at distance $z=\ell$. The two defects are parametrized as
\be
 J_a({\bf x }) = \delta(z)\,,\quad\quad 
 J_b({\bf x }) =  \delta^{d-2} ({\bf x}_\parallel)\delta(z-\ell)\,.
 \label{eq:source_pointmembrane}
\ee
{The membrane is infinitely thin in contrast with the point-plate case of the previous section where the plate had a large width.  }
We choose that the deformation moves the pointlike defect along $z$ while the membrane stays in place, i.e. it is given by \eqref{eq:deformation_gen} where $\Lv$ is oriented along $z$.  
The deformation of the defect is then given by 
\be
\partial_\lambda J =-\xi_b L \delta^{d-2} ({\bf x}_\parallel) \partial_z \delta(z-\ell)\,. 
\ee

The quantum force is given by
\begin{eqnarray}
F(\ell) & =&  -\frac{1}{2} \int d^{d-1} \x \, \langle{\cal O}(x){\cal O}(x)\rangle_J
\, \partial_\lambda J_{b,\lambda}(\x) 
\\
& =& -\frac{\xi_b}{2}  \, \partial_z  
\langle{\cal O}(x^\alpha,z){\cal O}(x^\alpha,z)\rangle_J   \big|_{z\to \ell} \,.
\end{eqnarray}
Here $\langle{\cal O}(x^\alpha,z){\cal O}(x^\alpha,z)\rangle_J  $ is the CFT 2-point function dressed by  the membrane at $z=0$. This correlator  is computed in Eq.\,\eqref{eq:2point_membrane}. 
Going to  position-momentum space, we have
\be G_J(p;z,z) = G_0(p)+
G^2(p;0,z)\frac{\xi_a}{1 -\xi_a G_0(p)}\,. \label{eq:G_membrane_coincident} \ee
The quantum force is then expressed as
\begin{eqnarray}
F(\ell)  & = &
-\frac{i\xi_b}{2} \int \frac{d^{d-1}p}{(2\pi)^{d-1}} \, \partial_z  G_J(p;z,z)  \big|_{z\to \ell} \\
& = & 
-\frac{i}{2} \int \frac{d^{d-1}p}{(2\pi)^{d-1}} \, \frac{\xi_a\xi_b}{1 -\xi_a G_0(p)}   \partial_z\left(  G^2(p;0,z)\right)_{z\to \ell} \label{eq:F_pointmembrane1}
\end{eqnarray}
where the first term from Eq.\,\eqref{eq:G_membrane_coincident} does not contribute since it is constant in $z$. $G_0(p)$ is defined in Eq.\,\eqref{eq:G0}.

The derivative piece takes a simple form
\be
 \partial_z\left(  G^2(p;0,z)\right)_{z\to \ell} = -
\frac{16\pi^{d-1}\ell}{\Gamma^2(\Delta)}\left(\frac{4\ell^2}{-p^2}\right)^{\frac{d-2-2\Delta}{2}} K_{\frac{1-d}{2}+\Delta}\left(\sqrt{-p^2}\ell\right) K_{\frac{3-d}{2}+\Delta}\left(\sqrt{-p^2}\ell \right) 
\ee
One may notice it is proportional to the product of two correlators with dimension $\Delta$ and $\Delta+1$. 

We can identify the potential directly from the line \eqref{eq:F_pointmembrane1}, where the $\partial_z$ derivative is equivalent to $\partial_\ell$.  
We rotate to Euclidian momentum $q^M$ and use spherical coordinates. 
We find the general result
\be
V(\ell) = - \frac{\pi^{\frac{d-1}{2}}}{2^{d-3}\Gamma(\frac{d-1}{2})\Gamma^2(\Delta)}\int dq q^{d-2} \, \frac{\xi_a\xi_b}{1 -\xi_b G_0(q)}   
\left(\frac{2\ell}{q}\right)^{d-1-2\Delta} K^2_{\frac{1-d}{2}+\Delta}\left(q\ell\right) \,.
\ee
\be
F(\ell)   = - \frac{\pi^{\frac{d-1}{2}}\ell}{2^{d-5}\Gamma(\frac{d-1}{2})\Gamma^2(\Delta)}\int dq q^{d-2} \, \frac{\xi_a\xi_b}{1 -\xi_b G_0(q)}   
\left(\frac{2\ell}{q}\right)^{d-2-2\Delta} K_{\frac{1-d}{2}+\Delta}\left(q\ell\right) K_{\frac{3-d}{2}+\Delta}\left(q\ell \right) \,.
\ee

We can evaluate the loop integral in both the Casimir-Polder regime $\xi_a G_0(p)\ll 1$ and the Casimir regime $\xi_a G_0(p)\gg 1$. 
We write the two Bessel functions using the representation Eq.\eqref{eq:Krep}, perform the loop momentum integral and then the $t$ and $t'$ integrals. The intermediate steps involve hypergeometric functions, but the final results are remarkably simple.

In the Casimir-Polder regime, we obtain the potential
\be
V(\ell)= - \frac{\pi^{(d-1)/2}\,\Gamma(2\Delta+\frac{1-d}{2})}{2\,\Gamma(2\Delta)}
  \frac{\xi_a\xi_b}{\ell^{4\Delta -d+1}}\,. 
\ee
It is attractive for any $\Delta$ satisfying the unitarity bound. 
For $d=3,4$  the Casimir-Polder force is
\be
F_{d=3}(\ell)   = -\frac{\pi \xi_a\xi_b}{\ell^{4\Delta-1}} \,,
\ee
\be
F_{d=4}(\ell)   = \frac{\pi^{3/2}\Gamma(2\Delta-\frac{1}{2})}{\Gamma(2\Delta)} \frac{\xi_a\xi_b}{\ell^{4\Delta-2}} \,.
\ee
For the free field in $d=4$,  including two factors of $a=\frac{1}{4\pi^2}$ to recover canonical normalization, we find $F(\ell)=-\frac{\xi_a\xi_b}{32\pi^2\ell^2}$. 
{Notice that this Casimir-Polder limit corresponds to a loop  between a point and  an infinitely thin membrane, it differs from the point-plate geometry of section \ref{se:CP_pointplate} where the width of the plate is large.}

In the Casimir regime, we obtain
 \be
 V(\ell) = - \frac{\sqrt{\pi} d\, \Gamma(d-1-\Delta)}{2^d\Gamma(1+\frac{d}{2})\Gamma(\frac{d-1-2\Delta}{2})} \frac{\xi_b}{\ell^{2\Delta}}  \,. 
  \ee
The potential depends only on $\xi_b $ and  is attractive for $\Delta$ in the interval of  interest, $\frac{d-2}{2}\leq\Delta<\frac{d-1}{2}$. 
For $d=3,4$ we obtain the forces
\be
F_{d=3}(\ell)   = \Delta(\Delta-1)\frac{\xi_b}{\ell^{2\Delta+1}} \,,
\ee
\be
F_{d=4}(\ell)   =  -  \frac{\sqrt{\pi}\Delta \, \Gamma(3-\Delta)}{4 \Gamma(\frac{3}{2}-\Delta)} \frac{\xi_b}{\ell^{2\Delta+1}} \,.
\ee
For the free field in $d=4$,  including one factor of $a=\frac{1}{4\pi^2}$ to recover canonical normalization, we find $F(\ell)=-\frac{\xi_b}{16\pi^2\ell^3}$. This reproduces exactly the Casimir force obtained from  the plate-point configuration taken in the Dirichlet limit
 computed in \cite{Brax:2022wrt}. This illustrates that, in the Casimir regime, only the boundary of the defect matters. 
{The $V(\ell)\propto \ell^{-2\Delta}$ dependence in the Casimir regime is  reminiscent of the fact that the ${\cal O}^2$ operator in a boundary CFT admits a vev with profile $\langle{\cal O}^2\rangle\propto z^{-2\Delta}$. 
}

\subsubsection{Membrane-Membrane CFT  Casimir}

We turn to the force between two membranes at $z=0$ and $z=\ell$. The two defects are parametrized as
\be
 J_a({\bf x }) = \delta(z)\,,\quad\quad 
 J_b({\bf x }) =  \delta(z-\ell)\,.
 \label{eq:source_membranemembrane}
\ee
The deformation of the defect is given by 
\be
\partial_\lambda J =-\xi_b L  \partial_z \delta(z-\ell)\,. 
\ee
Following the same steps as in the previous subsection, we arrive at the quantum pressure
\be
\frac{F(\ell)}{S_{d-2}}   = 
-\frac{i\xi_b}{2} \int \frac{d^{d-1}p}{(2\pi)^{d-1}} \, \partial_z  G_J(p;z,z)  \big|_{z\to \ell} \label{eq:Pressure_membranes1}
\ee
with $S_{d-2}=\int d^{d-2}\x_\parallel$. 
The 2-point correlator in the presence of two membranes is computed in Eq.\,\eqref{eq:dressed_two_membranes}. 

Unlike in the other cases previously treated, it is not possible to identify a potential directly from Eq.\,\eqref{eq:Pressure_membranes1}. This is due to the fact that the $\partial_z$ derivative cannot be traded for a derivative in $\ell$, because the dressed 2-point correlator depends nontrivially on $\ell$,  see Eq.\,\eqref{eq:dressed_two_membranes}.  Rather, we first compute the force, which is the fundamental quantity, then one may optionally infer a potential from it. 

Our focus is on the Casimir limit, which amounts to taking large $\xi_{a,b}$. Notice that one cannot use in Eq.\,\eqref{eq:Pressure_membranes1} the Dirichlet limit of \eqref{eq:G_Dir}, for which $\xi_{a,b}=\infty$.  This would lead to an indefinite $0\times \infty$ form in Eq.\,\eqref{eq:Pressure_membranes1}. Instead, one should compute the expansion of $G_J$ for large but finite $\xi_{a,b}$. The self-consistency of the quantum work formalism ensures that this expansion and the $\xi_b$ factor in  Eq.\,\eqref{eq:Pressure_membranes1} will conspire to give a finite result for the pressure.

We find 
\be
\partial_z  G_J(p;z,z)  \big|_{z\to \ell} = \frac{1}{\xi_b}\frac{\partial_z(G^2(0,z))_{z\to\ell}}{G^2_0(p)-G^2(p;0,\ell)} +O\left(\frac{1}{\xi_a},\frac{1}{\xi_b^2}\right)
\ee
To obtain this result, we use that $\partial_z G(p;z,z')|_{z'\to z}=0$ by symmetry. This sets to zero the would-be leading  term $\xi_{a,b}^0$. As a result the $1/\xi_b$ term is the leading one.

The quantum pressure between the membranes is then 
\be
\frac{F(\ell)}{S_{d-2}}   = \frac{\pi^{\frac{d-1}{2}}}{2^{d-1}\Gamma(\frac{d-1}{2})} \int dq q^{d-2} \frac{\partial_z(G^2(q;0,z))_{z\to\ell}}{G^2_0(q)-G^2(q;0,\ell)} \,.\label{eq:Pressure_membranes2}
\ee
Using the identity
\be
\partial_z\log\left(G(q;0,z)\right)\big|_{z\to\ell} = -\frac{q K_{\Delta-\frac{d}{2}+\frac{3}{2}}(q \ell)}{K_{\Delta-\frac{d}{2}+\frac{1}{2}}(q \ell)}
\ee
we find the final form
\be
\frac{F(\ell)}{S_{d-2}}   = \frac{\pi^{\frac{d-1}{2}}}{2^{d-5-2\Delta}\Gamma(\frac{d-1}{2})} \int dq q^{d-1} \frac{ (q\ell)^{2d}K_{\Delta-\frac{d}{2}+\frac{3}{2}}(q \ell)K_{\Delta-\frac{d}{2}+\frac{1}{2}}(q \ell)}
{ 2^{3+2\Delta} (q\ell)^{2d}K^2_{\Delta-\frac{d}{2}+\frac{1}{2}}(q \ell) 
- 2^d (q\ell)^{2\Delta+2d+1}\Gamma^2(\frac{d-1-2\Delta}{2})
} \,.\label{eq:Pressure_membranes3}
\ee

As a sanity check, for a free field ($\Delta=\frac{d-2}{2}$) we  recover exactly the well-known Casimir pressure between two membranes in any dimension \cite{Ambjorn:1981xw}.
The quantum pressure between the membranes is negative on the $\frac{d-2}{2}<\Delta<\frac{d-1}{2}$ interval. It is independent on $\xi_{a,b}$ and scales as $
\frac{F(\ell)}{S_{d-2}} \propto \frac{1}{\ell^d} $
as can be seen from Eq.\,\eqref{eq:Pressure_membranes3}.

We see that the Casimir regime displays a sense of universality. In the Casimir regime, the pressure does not depend on the strength of the double-trace couplings $\xi_{a,b}$. The pressure scales as $\ell^{-d}$ just like for 
a weakly coupled CFT, this scaling is dictated by the geometry of the problem. The sign of the force is also fixed, see next section. The only non-trivial data is the strength of the force. One can check via numerical integration that the strength of the force does depend on $\Delta$. Hence, in spite of the screening,  information about the double-trace nature of the boundary still remains encoded in the overall coefficient of the pressure.

\subsection{Monotonicity from Consistency}

In the above results, it may seem that the $\xi_{a,b}$ coefficients can be arbitrary real numbers, such that the $\xi_a\xi_b$ product can get both signs and thus that some of the  forces may be either attractive or repulsive. We show that this is not the case.

From section \ref{se:CasimirPolder}, it is clear that the quantum force between any two bodies in the Casimir-Polder (i.e. UV) regime  has the sign of $-\xi_a\xi_b$, i.e. it is attractive (repulsive) if $\xi_a\xi_b>0$ $(\xi_a\xi_b<0)$. 
On the other hand, we have found in section \ref{se:CasimirCasimir} that the force between two membranes in the Casimir (i.e. IR) regime is negative independently of $\xi_{a,b}$. 
None of these observations in itself constrains $\xi_{a,b}$, but one may note that if $\xi_a \xi_b<0$, then the force would have to flip sign in the transition from Casimir-Polder to Casimir. To understand whether such a behavior is allowed, we need to consider the exact fomulas that interpolates between the UV and IR regimes.

First consider the point-membrane configuration given in Eq.\,\eqref{eq:F_pointmembrane1}. 
We focus on the dressed 2-point correlator shown Eq.\,\eqref{eq:G_membrane_coincident}. 
For $\frac{d-2}{2}\leq \Delta<\frac{d-1}{2}$, we have  $G_0(p)\in \mathbb{R}_-$ for spacelike or Euclidian momentum. 
This implies that if $\xi_a<0$, then the dressed correlator features a pole 
at real negative $p^2$. This is a tachyon, whose mass is
\be
m_{\rm tachyon}^2= - 4 \left(-\frac{\Gamma(\Delta)}{\pi^{\frac{d-1}{2}}\Gamma(\frac{d-1-2\Delta}{2})\,\xi_a}\right)^{\frac{2}{2\Delta-d+1}}\,. \label{eq:tachyon}
\ee
 The presence of the tachyon pole has a very concrete consequence:  having $\xi_a<0$ would make  the loop integral in Eq.\,\eqref{eq:F_pointmembrane1} divergent. Since the force must be finite, this possibility is ruled out. Therefore $\xi_a$ must be positive. 

As similar analysis can be performed in the membrane-membrane configuration. For example, the same  tachyon mass Eq.\,\eqref{eq:tachyon}  shows up if one let one of the $\xi_i$ be small.  The tachyon pole also exists if e.g. $\xi_a=\xi_b$ , in that case the tachyon mass receives a $\ell$-dependent correction from the $G^2(p;0,\ell)$ term. We conclude that again $\xi_{a}$ and $\xi_b$ must be positive. 

Having $\xi_{a,b}>0$ implies that the force does not change sign for any value of the separation $\ell$. In other words, the absence of the tachyon is tied to the  potential being monotonic.

A similar reasoning involving a tachyon has been done for a double-trace deformation occupying all spacetime in \cite{Porrati:2016lzr}. In this reference the existence of the tachyon for $\xi<0$ is understood as  an obstruction to the RG flow --- while for $\xi>0$ there is no obstruction. Our argument here can be seen as an analogous version of this obstruction  statement for a situation where the double-trace deformation is localized on a membrane. The said obstruction appears concretely when computing the quantum force.

We briefly mention that  in the $\frac{d-1}{2}<\Delta<\frac{d}{2}$ case, the sign of the $\Gamma(\frac{d-1-2\Delta}{2})$ factor that appears in Eq.\,\eqref{eq:tachyon}  becomes positive. Applying the above chain of arguments  would then imply that  $\xi$ should be negative in this range of $\Delta$. However, as pointed out in section \ref{se:membrane}, the computations likely cannot be trusted in this domain --- extra effort would be needed to appropriately treat the divergent piece in $G_0$ (see eq.\,\eqref{eq:2point_smallz12}). 

Finally, let the  support of the defect, $J(\x)$,  be interpreted  not just as an abstract distribution but as a physical density of matter. At the level of the Lagrangian this is easily written covariantly  by coupling ${\cal O}^2$ to the trace of the stress-energy tensor $T_{\mu}^\mu$, with 
\be
{\cal L}=-\frac{\xi}{2m}{\cal O}^2 T_{\mu}^\mu(x)\,
\ee
with $m$ the mass of the matter particle.
In the presence of non-relativistic, static matter, we simply have $T_{\mu}^\mu(x)=\rho(\x)=m n(\x)$ with $n(\x)$ the number density. 
 Then the generic $\xi_{a,b}$ parameters that we have been using for each defects
are related to a single fundamental coupling   $\xi_i=n_i \xi$. 
In that view, any of the above arguments that constrain some of the $\xi_a$ to be positive  implies that $\xi>0$. It then follows that the $\xi_i$ of any defects are positive, therefore the potential between any two defects is monotonic.
In other words, under the condition that  $J$ be interpretable as  a physical density,
the quantum force between any two defects is attractive at any value of their separation.\,\footnote{  
The notion of $J$ being interpretable as  a physical density is also needed to ensure finiteness of the quantum work \cite{Brax:2022wrt}. }

\label{se:positivity}

\subsection{Critical Casimir Forces}

We briefly connect our results to critical Casimir forces.
We simply present the scalings predicted from our double trace model in the geometries considered in Secs.\,\ref{se:CasimirPolder},\,\ref{se:Casimir}. 
For thermal fluctuations at criticality, the relevant quantity is $\beta_c \delta F$ with $\beta_c=1/T_c$ where $\delta F$ is the geometry-dependent term of the free energy.  $\beta_c \delta F$ has vanishing mass dmension.

In the Euclidian field theory, the coupling of the double trace operator to the source is $\frac{1}{2}\int d^dx_E \xi {\cal O}^2(x_E) J(x_E)$. The $\xi$ coupling has  a scaling dimension $[\xi] = d -[J]-2\Delta$. 
The behavior of the forces follows by dimensional analysis. 


The free energy in the short distance limit gives non-retarded van der Waals forces. 
In the point-point, plate-point and plate-plate  geometries we obtain 
\be
\beta_c \delta F|_{\rm pt-pt} \propto \frac{\xi_a\xi_b }{\ell^{4\Delta}}\,, \quad 
\beta_c \delta F|_{\rm plate-pt} \propto \frac{\xi_a\xi_b }{\ell^{4\Delta-d}}\,, \quad 
\beta_c \delta F|_{\rm plate-plate} \propto \frac{\xi_a\xi_b }{\ell^{4\Delta-2d}}\,.
\ee
In the long distance limit, this gives Casimir-type forces. The membrane-point and membrane-membrane results are 
\be
\beta_c \delta F|_{\rm memb-pt} \propto \frac{\xi}{\ell^{2\Delta}}  \,,\quad\quad  
S^{-1}_{d-1} \beta_c \delta F|_{\rm memb-memb} \propto   \frac{1}{\ell^{d-1}}  \,. 
\ee
{  $\xi$ is the coupling to the pointlike defect.
The couplings to  the membranes do not appear in the Casimir limit. 
In the membrane-membrane case we give the the free energy  per units of  area of the membrane, $S_{d-1} $.}
The point-point and membrane-point results match predictions made from limits of the sphere-sphere geometry in the critical Ising model\,\cite{PhysRevLett.74.3189,PhysRevB.51.13717,PhysRevLett.81.1885}.     


\section{Summary}
\label{se:summary}

We explore the quantum forces occurring between the defects and/or boundaries of conformal field theories.  While  defect CFTs  are often investigated formally, our approach here is more concrete. Since such CFTs do exist in the laboratory, our focus is to predict phenomena  that may, at least in principle, be experimentally observed. 
Our computations only require basic notions of CFT and a solid formalism to derive quantum forces in arbitrary situations.

{Defects and boundaries in the real world are not perfect, in the sense that no real-world material  can truncate the spatial support of a field theory fluctuating at all wavelengths.} Inspired by models used in weakly coupled QFT, we propose to model imperfect defects of CFTs as localized relevant double-trace operators. This idea is  nicely supported by the $\Delta_-$ branch of the AdS/CFT correspondence, in which case the defects are identified as mass terms localized on the (regularized) boundary of the Poincaré patch.

In order to compute quantum forces, we need to know
 the 2-point CFT correlators in the presence of such ``double-trace'' defects. {Assuming large $N$,} this is described by a Born series that dresses the CFT correlator with insertions of the defect. 
We first make clear that the CFT correlators get repelled from the defects in the infrared regime. Asymptotically in the IR, the CFT satisfies a Dirichlet condition on the boundary of the defect. In this limit the interior of the defect becomes irrelevant.

The archetype of an extended defect is the codimension-one hyperplane, i.e. the membrane. In the presence of a membrane we restrict the conformal dimension to $\frac{d-2}{2}\leq \Delta<\frac{d-1}{2}$ to avoid dealing with a divergence in the membrane-to-membrane correlator. A careful analysis of the $\Delta>\frac{d-1}{2}$ case remains to be done.

 We compute the 2-point correlator in the presence of two parallel membranes
and investigate some if its  features. We find that the CFT between the membranes develops a sequence of poles away from the real axis, that should be understood as a set of resonances, or collective excitations, of the CFT constituents. In the near-free limit, these resonances are narrow with decay rate depending only on the separation between the two membranes and on the dimension of the double-trace operator. 
It would be interesting to  study further the properties of these resonances, including their interactions.

We then explore the quantum forces between pointlike and/or planar double-trace defects in the asymptotic Casimir-Polder and Casimir regimes.  
The Casimir-Polder regime typically appears at short separation i.e. in the UV, when the first term of the Born series are leading. 
The CFT Casimir-Polder force between a pointlike defect and either another  pointlike defect, a membrane, or an infinite plate, is respectively proportional to $1/\ell^{4\Delta-2}$, $1/\ell^{4\Delta-d+2}$, $1/\ell^{4\Delta-d+1}$. The  force between two infinite plates is in $1/\ell^{4\Delta-d}$.

The Casimir regime appears at large separation, i.e. in the IR, when the Born series must be resummed. The Casimir force between a point and a membrane goes as $1/\ell^{2\Delta+1}$, while the pressure between two membranes goes as $1/\ell^{d}$.  The membrane-membrane quantum pressure has, in a sense, a universal behavior analogous to the one induced from free fields. 
However, information about the double-trace nature of the boundary still  remains in the overall coefficient of the force,  which is $\Delta$-dependent.

In membrane configurations, we show that the sign of the double-trace operator is constrained in order for the potential to be well-defined at any distance. This is tied to requiring the absence of a tachyon in the spectrum of the two-point correlator. In turn, this constraint  guarantees that the potential is 
monotonic. Assuming that the support of the defects can be interpreted as a physical matter distribution --- an assumption that is also needed to ensure finiteness of the quantum work, one concludes that the potential between any two defects is monotonic. Hence, the quantum forces between any two  double-trace defects are attractive at any distance. 

{It would be interesting to determine real world  systems {--- either quantum or critical ---}  for which the defects and boundaries may, at least approximately, be described by  double-trace deformations. It would also be interesting to devise laboratory experiments that can test some of the phenomena predicted in this paper. 
The exploration of these possibilities is left for future work.
}

\appendix

\section{Two-point Correlator in Mixed Space }
\label{app:twopoint}

The Schwinger parametrization is 
\be
\frac{1}{x^{2\Delta}_{12}} =\frac{1}{\Gamma(\Delta)}\int^\infty_0 \frac{dt}{t}t^\Delta e^{-t x^2_{12}}\,. \label{eq:Schwing}
\ee
We use it to compute the Fourier transform
\begin{align}
\llangle  {\cal O}(p,z_1)  {\cal O}(-p,z_2) \rrangle & = \int d^{d-1}y_{12}\,e^{iy_{12}\cdot p}\frac{1}{x^{2\Delta}_{12}} \\
& =
 \frac{1}{\Gamma(\Delta)}\int^\infty_0  \frac{dt}{t}t^{\Delta -tz_{12}^2}
 \int d^{d-1}y_{12}\,e^{iy_{12}\cdot p}
 e^{-t y^2_{12}} 
 \\
& = -i
  \frac{\pi^{\frac{d-1}{2}}}{\Gamma(\Delta)}\int^\infty_0  \frac{dt}{t}t^{\Delta-\frac{d-1}{2}} 
  e^{-t z^2_{12}+\frac{p^2}{4t}} \,.  \label{eq:TFcorr3}
\end{align}
In the second line the time integral is evaluated upon Wick rotation to Euclidian space, $y^0_{12}=-i y^{0,E}_{12} $, that makes appear the overall $-i$ factor.
In the last line we recognize the integral representation of the Bessel $K$ given in Eq.\,\eqref{eq:Krep}, that we can put in the form 
\be
 \int_0^\infty \frac{dt}{t} t^{\Delta-\frac{d-1}{2}} e^{-tz^2-\frac{q^2}{4t}} = 
 2 \left(\frac{q}{ 2 z }\right)^{\Delta-\frac{d-1}{2}} K_{\Delta-\frac{d-1}{2}}(  qz)\,. \label{eq:TFcorr4}
\ee
We remind that $K_\alpha(z)=K_{-\alpha}(z)$. 
Identifying Eq.\,\eqref{eq:TFcorr4} in Eq.\,\eqref{eq:TFcorr3}, we obtain the momentum-position representation of the two-point correlator presented in Eq.\,\eqref{eq:2pt_mixed}.

\section{ Computation of the Decay Widths }

\label{app:widths}

Consider the denominator of Eq.\,\eqref{eq:G_Dir}, 
\be
D(p)=G^2_0(p)-G^2(p;0,L)\,. 
\ee
For $\Delta=\frac{d-2}{2}$, we have 
\be
D(p)= \frac{4\pi^d}{\Gamma^2(\frac{d-2}{2})}\frac{1-e^{-2L\sqrt{-p^2}}}{p^2}\,. \label{eq:Den_free}
\ee
In that case, $D(p)$ has a set of zeros on the real line, $D(m^{\rm free}_n)=0$, at the values 
$p=m^{\rm free}_n\equiv\frac{n\pi}{L}$, $n\in\mathbb{N}_{/0}$. 
These are the familiar modes   of the free field confined in a $[0,L]$ Dirichlet interval. 

For $\Delta$ close to the free field dimension, we can expand the denominator in $\epsilon=\Delta-\frac{d-2}{2}$. This produces a small correction to  Eq.\,\eqref{eq:Den_free}. 
We get
\be
D(p)\approx  \frac{4\pi^d}{\Gamma^2(\frac{d-2}{2})} \left(\frac{\sqrt{-p^2}}{L}\right)^\epsilon \frac{\left( L \sqrt{-p^2}\right)^{2\epsilon}-e^{-2L\sqrt{-p^2}}}{p^2} \label{eq:Den}.
\ee
By continuity the poles are given by $m_n=m_n^{\rm free}+
\epsilon \delta_{n,r}+
i\epsilon \delta_{n,i} +O(\epsilon^2)$.

We assume that the imaginary part of the $\delta$ correction is negative, $\delta_{n,i}<0$.
Plugging this form into Eq.\,\eqref{eq:Den} and expanding in $\epsilon$ determines the $\delta$ corrections. 
We find 
\be
\delta_{n,r}= - \epsilon \log\left(\frac{n\pi}{L}\right)  \,,
\ee
\be
\delta_{n,i}= - \epsilon \frac{\pi}{2 L} \,.
\ee
We have thus $\delta_{n,i}<0$, consistent with our hypothesis. These poles describe narrow resonances. 
In particle physics the imaginary part is usually written as 
\be
\delta_{n,i}\equiv -\frac{\Gamma_n}{2}\,,
\ee
where $\Gamma_n\ll m_n$ is the decay rate i.e. the width of the resonance. This leads to the formula Eq.\,\eqref{eq:widths}.

\bibliographystyle{JHEP}
\bibliography{biblio}

\end{document}